\documentclass[english,onecolumn]{IEEEtran}
\usepackage[T1]{fontenc}
\usepackage[latin9]{inputenc}
\usepackage{amsmath}
\usepackage{amsthm}
\usepackage{amssymb}
\usepackage{graphicx}
\usepackage{wasysym}
\usepackage{esint}

\makeatletter

\providecommand{\tabularnewline}{\\}

\theoremstyle{plain}
\newtheorem{thm}{\protect\theoremname}
\theoremstyle{remark}
\newtheorem{rem}[thm]{\protect\remarkname}
\theoremstyle{plain}
\newtheorem{cor}[thm]{\protect\corollaryname}
\theoremstyle{plain}
\newtheorem{conjecture}[thm]{\protect\conjecturename}

\usepackage{mathrsfs}
\usepackage{algorithm,algpseudocode}

\usepackage{colortbl}
\definecolor{lightgray}{rgb}{0.9,0.9,0.9}
\definecolor{lightred}{rgb}{1,0.8,0.8}
\definecolor{lightgreen}{rgb}{0.6,1,0.6}
\definecolor{lightyellow}{rgb}{1,1,0.5}
\definecolor{lightgrey}{rgb}{0.8,0.8,0.8}

\allowdisplaybreaks[1]

\makeatother

\usepackage{babel}
\providecommand{\conjecturename}{Conjecture}
\providecommand{\corollaryname}{Corollary}
\providecommand{\remarkname}{Remark}
\providecommand{\theoremname}{Theorem}

\begin{document}
\title{The Undecidability of Network Coding with some Fixed-Size Messages
and Edges}
\author{Cheuk Ting Li\\
Department of Information Engineering, The Chinese University of Hong
Kong\\
Email: ctli@ie.cuhk.edu.hk}
\maketitle
\begin{abstract}
We consider a network coding setting where some of the messages and
edges have fixed alphabet sizes, that do not change when we increase
the common alphabet size of the rest of the messages and edges. We
prove that the problem of deciding whether such network admits a coding
scheme is undecidable. This can be considered as a partial solution
to the conjecture that network coding (without fixed-size messages/edges)
is undecidable. The proof, which makes heavy use of analogies with
digital circuits, is essentially constructing a digital circuit of
logic gates and flip-flops within a network coding model that is capable
of simulating an arbitrary Turing machine.
\end{abstract}


\medskip{}

\section{Introduction}

Network coding \cite{ahlswede2000network,li2003linear} concerns the
setting where messages are transmitted through a network of nodes,
where each node is capable of performing encoding and decoding operations
(instead of being limited to routing operations). There is a wealth
of algorithms and computability results for network coding, e.g. \cite{jaggi2005polynomial,harvey2005deterministic,li2005achieving,ho2006random}.
However, the other side of the picture -- the hardness of network
coding -- is less clear. While linear network codes \cite{li2003linear,ho2006random}
are usually simpler to design and analyze, they are generally insufficient
in achieving the capacity \cite{dougherty2005insufficiency,chan2008dualities}.
It was shown by Rasala Lehman \cite{lehman2005network} that deciding
whether a network admits a coding scheme with a given alphabet size
is NP-hard. Langberg, Sprintson and Bruck \cite{langberg2006encoding}
showed that determining the minimum number of encoding nodes in a
multicast network is NP-hard. Langberg and Sprintson \cite{langberg2011hardness}
showed that approximating the capacity of a network is NP-hard. Refer
to \cite{lehman2004complexity,harvey2005deterministic,lehman2005networkmodel,yao2009network}
for more results on the hardness of network coding.

Nevertheless, the NP-hardness results do not completely settle the
question on the hardness of network coding. It is currently unknown
if network coding is even decidable, i.e., if there exists an algorithm
where, given the network as input, outputs whether the network admits
a coding scheme satisfying the decoding requirements \cite{lehman2005network,cannons2006network,dougherty2009undecidable,dougherty2011network,bassoli2013network,gomez2014network,kuhne2019representability}.
For arguments in favor of the decidability of network coding, Rasala
Lehman \cite{lehman2005network} noted that network coding would be
decidable if one can compute a finite upper bound on the alphabet
size given a network. If only routing is allowed, then the routing
capacity of the network is computable, as shown by Cannons et al.
\cite{cannons2006network}. For general network codes, computable
inner bounds (e.g. \cite{cannons2006network}) and computable outer
bounds (e.g. \cite{thakor2009network}) are known, though the computability
of the capacity is unknown.

For arguments against decidability, K{\"u}hne and Yashfe \cite{kuhne2019representability}
showed that the problem of finding whether a network admits a vector
linear network code is undecidable, by proving an undecidability result
about matroids and invoking the relationship between vector linear
network codes and matroids in \cite{el2010index}, though the relationship
for general network codes is less clear. Dougherty \cite{dougherty2009undecidable}
proposed a possible approach to prove undecidability via a reduction
from Rhodes' problem (the identity problem for finite groups), which
is conjectured to be undecidable \cite{albert1992undecidability}
(though there are holes in the arguments in \cite{dougherty2009undecidable};
also see \cite{dougherty2011network}). 

A closely related line of research is the characterization of the
the entropic region $\Gamma_{n}^{*}$ and the almost-entropic region
$\overline{\Gamma_{n}^{*}}$ (the closure of $\Gamma_{n}^{*}$) \cite{zhang1997non,yeung1997framework,zhang1998characterization}.
Their relation with network coding was elucidated in \cite{yeung2008information,chan2008dualities,yan2012implicit}.
Non-Shannon-type inequalities, which are bounds on $\Gamma_{n}^{*}$,
were studied in \cite{zhang1997non,zhang1998characterization,makarychev2002new,dougherty2006six,matus2007infinitely,xu2008projection,dougherty2011non}.
It was shown by Li \cite{li2021undecidability} that the problem of
deciding whether $\Gamma_{n}^{*}$ intersects a given affine subspace
is undecidable. See \cite{gomez2014network,gomez2017defining,gomez2018theory,khamis2020decision,li2021first}
for other works related to the decidability or undecidability of problems
regarding $\Gamma_{n}^{*}$.

In this paper, we consider a network coding setting, which we call
the \emph{partially fixed-size network}, where some of the messages
and edges have given fixed alphabet sizes. All other messages and
edges have a common alphabet size $k$. We show that the problem of
deciding whether there exists $k$ such that this network admits a
coding scheme is undecidable. This is the first undecidability result
about general (linear or nonlinear) network codes. Note that if none
of the messages and edges have fixed sizes, then the problem becomes
the original network coding problem, and its decidability is unknown
\cite{lehman2005network,dougherty2009undecidable}. On the other hand,
if all messages and edges have fixed sizes, then the problem is clearly
decidable by the arguments in \cite{lehman2005network} (since one
can enumerate all encoding and decoding functions). It is perhaps
surprising that combining these two cases makes the problem undecidable.

Due to the equivalence between network coding and index coding \cite{bar2011index,lubetzky2009nonlinear}
proved in \cite{el2010index,effros2015equivalence}, we can also show
the undecidability of an index coding problem where the sizes of some
messages are fixed.

The undecidability is proved via a reduction from the periodic tiling
problem \cite{wang1961proving,berger1966undecidability,gurevich1972remarks,mazoyer1999global}.
This is the same strategy employed in the proof of the the undecidability
of conditional affine information inequalities and conditional independence
implication with a binary constraint by Li \cite{li2021undecidability}.
The proof in this paper shares many similarities with the proof in
\cite{li2021undecidability}. Nevertheless, the network coding setting
is significantly more restrictive than the information inequality
and the conditional independence setting in \cite{li2021undecidability}
(e.g. it is impossible to enforce that the signals along two edges
are independent in network coding), making the proof in this paper
considerably more challenging.\footnote{We remark that a natural approach to prove the undecidability of partially
fixed-size network is to invoke the result that deciding whether $\Gamma_{n}^{*}$
intersects a given affine subspace is undecidable \cite{li2021undecidability},
and the duality result between $\overline{\Gamma_{n}^{*}}$ and network
coding \cite{chan2008dualities}. This approach does not work since
\cite{li2021undecidability} only shows the undecidability of the
intersection problem, not the problem of deciding whether a given
vector is in $\overline{\Gamma_{n}^{*}}$. Nevertheless, \cite{chan2008dualities}
requires a specific vector $\mathbf{h}$ in order to design the capacities
of the edges in the network that is asymptotically solvable if and
only if $\mathbf{h}\in\overline{\Gamma_{n}^{*}}$.}

In the proof, we use several analogies with digital circuits, such
as XOR gates, tristate buffers, switches, flip-flops and memory arrays.
Using these components, we construct the tiles of the periodic tiling
problem, which in turn can be used to construct a Turing machine\footnote{The undecidability of the periodic tiling problem \cite{gurevich1972remarks,mazoyer1999global}
was proved via a reduction from the halting problem \cite{turing1937computable}.}. Therefore, the proof can be regarded as constructing a Turing machine
within a network coding model using digital circuit components. This
approach is perhaps unexpected. While hardware implementation of network
coding using digital circuits has been studied (e.g. \cite{yoon2010fpga}),
it is quite unusual to liken the communication network itself to a
digital circuit (with ``wires'' being the communication links, and
``logic gates'' being composed of nodes in the network).

A consequence of our result is that there is an explicit construction
of a partially fixed-size network, where the non-existence of a coding
scheme is unprovable in ZFC (assuming ZFC is consistent). This is
because there is a Turing machine such that whether it halts is independent
of ZFC, assuming ZFC is consistent \cite{michel2009busy,yedidia2016relatively}.

\medskip{}

\subsection*{Notations}

We write $\mathbb{N}_{0}=\{0,1,\ldots\}$, $\mathbb{N}_{+}=\{1,2,\ldots\}$,
$[a..b]:=\mathbb{Z}\cap[a,b]$, $[n]:=\{1,\ldots,n\}$, $X_{a}^{b}:=(X_{a},X_{a+1},\ldots,X_{b})$,
$X^{n}:=X_{1}^{n}$. For a finite set $S\subseteq\mathbb{N}_{+}$,
we write $X_{S}:=(X_{a_{1}},\ldots,X_{a_{k}})$, where $a_{1},\ldots,a_{k}$
are the elements of $S$ in ascending order. For a random variable
$X$, write $\mathcal{X}$ for its support, and $|\mathcal{X}|$ for
its alphabet size. For a probability mass function $p_{X}$, write
$\mathrm{supp}(p_{X})=\{x\in\mathcal{X}:\,p_{X}(x)>0\}$ for its support.
For a countable set $S$, write $\mathcal{P}(S)$ for the set of probability
measures over the sample space $S$. The indicator of an event or
statement $E$ is written as $\mathbf{1}\{E\}\in\{0,1\}$.

\medskip{}

\section{Problem Formulation}

A partially fixed-size network is represented by a directed acyclic
graph $(V,E)$. Write the set of in-neighbors and out-neighbors of
$v\in V$ as $N_{\mathrm{in}}(v)\subseteq V$ and $N_{\mathrm{out}}(v)\subseteq V$
respectively. There are $l$ independent messages $M_{1},\ldots,M_{l}$.
The alphabet size of the message $M_{i}$ is $s_{M}(i)\in\mathbb{N}_{0}$,
where $s_{M}(i)=0$ is a special value which means that $M_{i}$ has
size $k$, where $k\in\mathbb{N}_{+}$ is the \emph{common default
alphabet size} of the network. We have 
\[
M_{i}\sim\mathrm{Unif}[0\,..\,s_{M}(i)+k\cdot\mathbf{1}\{s_{M}(i)=0\}-1],
\]
i.e., the range of $M_{i}$ is $0,\ldots,s_{M}(i)-1$ if $s_{M}(i)>0$,
and $0,\ldots,k-1$ if $s_{M}(i)=0$. If $s_{M}(i)=0$, we call $M_{i}$
a \emph{default-size message}. Otherwise, we call $M_{i}$ a \emph{fixed-size
message}. Node $v\in V$ has access to messages $M_{A_{v}}$, and
wants to decode the messages $M_{B_{v}}$, where $A_{v},B_{v}\subseteq[l]$. 

Let the signal transmitted along edge $(u,v)\in E$ be $X_{u,v}\in\mathbb{N}_{0}$.
Each edge $(u,v)\in E$ has size $s_{E}(u,v)\in\mathbb{N}_{0}$, where
$s_{E}(u,v)=0$ is a special value with the same meaning as $s_{M}(i)=0$.
If $s_{E}(u,v)=0$, we call $(u,v)$ a \emph{default-size edge}. Otherwise,
we call $(u,v)$ a \emph{fixed-size edge}. Let the encoding function
for edge $(u,v)$ be 
\[
f_{u,v}:\,\mathbb{N}_{0}^{|A_{u}|+|N_{\mathrm{in}}(u)|}\to[0\,..\,s_{E}(u,v)+k\cdot\mathbf{1}\{s_{E}(u,v)=0\}-1].
\]
We have 
\[
X_{u,v}=f_{u,v}(M_{A_{u}},\{X_{t,u}\}_{t\in N_{\mathrm{in}}(u)}).
\]
The decoding function at node $v$ is $g_{v}:\,\mathbb{N}_{0}^{|N_{\mathrm{in}}(v)|}\to\mathbb{N}_{0}^{|B_{v}|}$.
The encoding and decoding functions do not need to be linear. The
decoding constraint is that $g_{v}(\{X_{u,v}\}_{u\in N_{\mathrm{in}}(v)})=M_{B_{v}}$
almost surely for all $v\in V$. 

We call the network $(V,E,\{A_{v}\},\{B_{v}\},s_{M},s_{E})$ \emph{solvable}
if there exists $k\in\mathbb{N}_{+}$ and a coding scheme $(\{f_{u,v}\},\{g_{v}\})$
such that the decoding constraint is satisfied. In the following sections,
we will prove the following undecidability result.

\medskip{}

\begin{thm}
\label{thm:network}The following problem is undecidable: Given a
partially fixed-size network $(V,E,\{A_{v}\},\{B_{v}\},s_{M},s_{E})$,
decide whether it is solvable.
\end{thm}
\medskip{}

Note that many previous models of network coding (e.g. \cite{lehman2004complexity,lehman2005network,lehman2005networkmodel,dougherty2005insufficiency,langberg2011hardness})
assume the alphabet sizes for all messages and edges are the same\footnote{In previous work, the messages and signals along edges are often considered
to be vectors of elements in the alphabet, instead of scalars as in
this paper. If the vectors are of fixed lengths, we can split the
messages into multiple scalar parts, and introduce parallel edges,
so that each message and signal can be considered as a scalar element
in the alphabet.}. This corresponds to the case where we always have $s_{M}(i)=0$
and $s_{E}(u,v)=0$. The decidability of the solvability of a network
in this case is still open \cite{dougherty2009undecidable}. Another
case is where we always have $s_{M}(i)>0$ and $s_{E}(u,v)>0$. In
this case, the solvability is clearly decidable using the argument
in \cite{lehman2005network}, since we can simply enumerate all combinations
of $\{f_{u,v}\},\{g_{v}\}$ (which are functions with domains and
codomains of bounded sizes) \footnote{Actually it suffices to have $s_{E}(u,v)>0$ for all $(u,v)\in E$
in order to show decidability, since it is always better to choose
$k=1$ in this case. Also, it suffices to have $s_{M}(i)>0$ for all
$i$ in order to show decidability, since we can choose $k$ large
enough such that each edge $(u,v)$ with $s_{E}(u,v)=0$ can simply
forward the received signals at node $u$.}. It is perhaps surprising that combining these two cases makes the
solvability undecidable.

\medskip{}

\section{Gates and Checkers}

In this section, we prove the main result by constructing a class
of undecidable partially fixed-size networks. We will make heavy use
of analogies with digital circuits in the proof, e.g. XOR gates, tristate
buffers, switches and 2D memory organization.

A \emph{checker} for the condition $\mathcal{Q}\subseteq\mathcal{P}(\mathbb{N}_{0}^{n})$
(a set of $n$-dimensional probability distributions) is a subnetwork
of the communication network with $n$ inputs $X_{1},\ldots,X_{n}$,
with a purpose of checking that $p_{X^{n}}\in\mathcal{Q}$. For example,
the XOR checker described later has three inputs $M_{1},M_{2},Y\in\{0,1\}$,
and checks that $Y=M_{1}\oplus M_{2}$ (up to relabelling of the values
of $Y$, i.e., $Y=1-M_{1}\oplus M_{2}$ is also valid). A checker
does not have any output. A \emph{gate} for the condition $\mathcal{Q}\subseteq\mathcal{P}(\mathbb{N}_{0}^{n_{1}n_{2}})$
is a subnetwork with inputs $X_{1},\ldots,X_{n_{1}}$ and outputs
$Y_{1},\ldots,Y_{n_{2}}$, where $p_{X^{n_{1}},Y^{n_{2}}}\in\mathcal{Q}$
is guaranteed to be satisfied (note that a gate may also check certain
conditions on the inputs). For example, the XOR gate described later
has two inputs $M_{1},M_{2}\in\{0,1\}$, and an output $Y\in\{0,1\}$
which satisfies $Y=M_{1}\oplus M_{2}$ (up to relabelling). 

\begin{table}
\begin{centering}
\begin{tabular}{|c|c|c|}
\hline 
$\begin{array}{c}
\text{Network coding}\\
\text{component}
\end{array}$ & $\begin{array}{c}
\text{Digital circuit}\\
\text{analogy}
\end{array}$ & Definition\tabularnewline
\hline 
\hline 
Edge in network & Wire & \tabularnewline
\hline 
Broadcast node & Junction & \tabularnewline
\hline 
Source node / message & Power supply & \tabularnewline
\hline 
$\begin{array}{c}
\text{Message signal to}\\
\text{gate/checker}
\end{array}$ & $\begin{array}{c}
\text{Power supply pin}\\
\text{of a chip}
\end{array}$ & Section \ref{subsec:xor}\tabularnewline
\hline 
$\begin{array}{c}
\text{Condition signal to}\\
\text{gate/checker}
\end{array}$ & $\begin{array}{c}
\text{Ground pin}\\
\text{of a chip}
\end{array}$ & Section \ref{subsec:xor}\tabularnewline
\hline 
Butterfly network & XOR gate & Section \ref{subsec:xor}\tabularnewline
\hline 
Tristate buffer gate & Tristate buffer & Section \ref{subsec:tristate}\tabularnewline
\hline 
Switch & $\begin{array}{c}
\text{2\ensuremath{\times}2 crossbar switch /}\\
\text{Flip-flop}
\end{array}$ & Section \ref{subsec:switch}\tabularnewline
\hline 
Conditional switch & Memory chip & Section \ref{subsec:conditional_switch}\tabularnewline
\hline 
$\begin{array}{c}
\text{Conditional switch with}\\
\text{tori select signal}
\end{array}$ & 2D memory organization & Section \ref{subsec:2d}\tabularnewline
\hline 
\end{tabular}
\par\end{centering}
\smallskip{}

\caption{Various components and constructions in network coding used in this
paper, and their corresponding digital circuit analogues.}

\end{table}

\medskip{}

\subsection{XOR Gate and Checker\label{subsec:xor}}

We construct the gate and checker for the exclusive or (XOR) function.
We first state a simple fact about XOR: for $X_{1},\ldots,X_{n}\stackrel{iid}{\sim}\mathrm{Bern}(1/2)$
and $Y\in\{0,1\}$, we have $H(X_{i}|X_{[n]\backslash\{i\}},Y)=0$
for all $i\in[n]$ if and only if $Y=X_{1}\oplus\cdots\oplus X_{n}$
or $Y=1-X_{1}\oplus\cdots\oplus X_{n}$. We give the proof for the
sake of completeness. Since $H(X_{1}|X_{[n]\backslash\{1\}},Y)=0$,
$Y$ is a function of $X^{n}$ (otherwise assume $p_{Y|X^{n}}(\cdot|x^{n})$
is nondegenerate for some $x^{n}$, and then it is impossible to decode
$X_{1}$ given $Y$ and $X_{2}^{n}=x_{2}^{n}$). Let $Y=f(X^{n})$.
Since $H(X_{1}|X_{[n]\backslash\{1\}},Y)=0$, we have $f(x_{1},\ldots,x_{n})\neq f(1-x_{1},x_{2},\ldots,x_{n})$,
and hence $f(x_{1},\ldots,x_{n})=1-f(1-x_{1},x_{2},\ldots,x_{n})$.
By repeated use of this relation, we have $f(x_{1},\ldots,x_{n})=f(0,\ldots0)$
if $\sum_{i}x_{i}$ is even, $f(x_{1},\ldots,x_{n})=1-f(0,\ldots0)$
if $\sum_{i}x_{i}$ is odd. The result follows.

Given messages $M_{1},M_{2}\stackrel{iid}{\sim}\mathrm{Unif}[0..1]$
and random variable $Y\in\{0,1\}$, we can check whether $Y=M_{1}\oplus M_{2}$
(up to relabelling of the values of $Y$) by checking $H(M_{1}|Y,M_{2})=H(M_{2}|Y,M_{1})=0$.
We denote this condition as $\mathrm{XOR}(M_{1},M_{2},Y)$.

\begin{figure}
\begin{centering}
\includegraphics[scale=0.92]{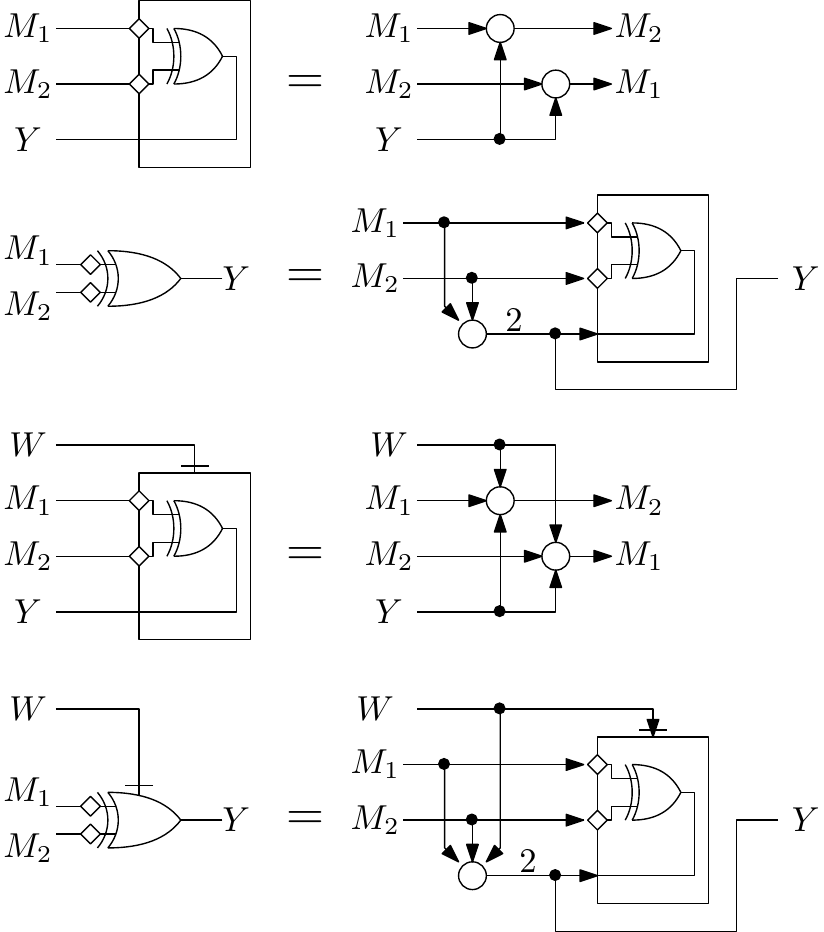}
\par\end{centering}
\caption{\label{fig:xor}Top: The XOR checker.  Middle 1: The XOR gate (identical
to the butterfly network \cite{ahlswede2000network}). Middle 2: The
conditional XOR checker. Bottom: The conditional XOR gate.}
\end{figure}

Figure \ref{fig:xor} shows the XOR checker. The label on an edge
is its size $s_{E}(u,v)$. Unlabelled edges are assumed to have unlimited
size (this is technically not allowed in the partially fixed-size
network, though we will circumvent this problem at the end of the
proof). The circles in the diagram (white circles and black dots)
represent nodes in the network. The black dots are broadcast nodes
(with in-degree one and with all outgoing edges having unlimited size)
where, without loss of generality, can be assumed to be sending its
input signal to each of its outgoing edge. The black dots can also
be considered as junctions in the circuit diagram (crossings without
dots are considered ``no contact''). The ``$\Circle\!\!\longrightarrow M_{1}$''
in the diagram means that the node ``$\Circle$'' has to decode
the message $M_{1}$. Therefore, the inputs $M_{1},M_{2}$ to the
XOR checker must be messages, though $Y$ can come from the intermediate
signals in the edges of the network. The small diamonds at the inputs
$M_{1},M_{2}$ indicate that these inputs must be messages (or combinations
of messages). They are called the \emph{message signals} to the checker.

Note that it is impossible to check for precise values of $Y$, and
it is impossible to distinguish between $Y=M_{1}\oplus M_{2}$ and
$Y=1-M_{1}\oplus M_{2}$. In general, a checker can only check a condition
on the random variables up to relabelling of those random variables. 

Technically, the XOR checker is a checker for $\mathcal{Q}$ that
is the set of joint distributions of $(M_{1},M_{2},Y)$, where $\mathcal{Q}$
satisfies that if $p_{M_{1},M_{2},Y}$ has $|\mathrm{supp}(p_{Y})|\le2$,
then $p_{M_{1},M_{2},Y}\in\mathcal{Q}$ if and only if $p_{M_{1},M_{2},Y}$
is a relabelling of the joint distribution $p_{\tilde{M}_{1},\tilde{M}_{2},\tilde{Y}}$
of $\tilde{M}_{1},\tilde{M}_{2}\stackrel{iid}{\sim}\mathrm{Bern}(1/2)$,
$\tilde{Y}=\tilde{M}_{1}\oplus\tilde{M}_{2}$.  Note that the checker
assumes that the support size of $Y$ is at most $2$, and fails to
check the XOR condition if this is not satisfied. The $\mathcal{Q}$
for all checkers and gates are defined via the same general pattern
(if the non-message inputs have the prescribed sizes, then the condition
is satisfied up to relabelling), and will be omitted in the remainder
of the paper.

The XOR gate (Figure \ref{fig:xor}) is constructed by creating an
output which is a binary function of $M_{1},M_{2}$ that must satisfy
the XOR condition enforced by the XOR checker. Note that the XOR gate
is identical to the butterfly network \cite{ahlswede2000network},
where the optimal code for binary alphabet uses the XOR operation.

We may also construct the conditional XOR checker, which has four
inputs $M_{1},M_{2},Y,W$ (where $M_{1},M_{2}\stackrel{iid}{\sim}\mathrm{Unif}[0..1]$
is independent of $W$; $W$ is called the \emph{condition signal}),
and checks that $Y=M_{1}\oplus M_{2}\oplus\eta_{W}$, where $\eta_{w}\in\{0,1\}$
for any $w$. Intuitively, conditional on any $W=w$, we have $Y=M_{1}\oplus M_{2}$
up to relabelling, where the relabelling can depend on $w$ (since
$Y$ is binary, a relabelling of $Y$ can be expressed as XOR with
$\eta_{w}$). In general, the conditional version of a checker for
$\mathcal{Q}\subseteq\mathcal{P}(\mathbb{N}_{0}^{n})$ conditioned
on $W$, which checks whether $p_{X^{n}|W=w}\in\mathcal{Q}$ for every
$w\in\mathcal{W}$ (i.e., checking whether the constraint of the checker
holds conditioned on any $W=w$), can be constructed by adding the
input $W$ to every non-broadcast node in the checker. To see why
this construction is valid, note that  for each $w$, if $p_{X^{n}|W=w}\in\mathcal{Q}$,
then there exists encoding and decoding functions for nodes in the
(unconditional) checker for $\mathcal{Q}$ which accepts $X^{n}\sim p_{X^{n}|W=w}$.
For an edge $(u,v)$, we can combine its encoding functions $f_{u,v,w}(Z)$
for different $w$'s together to form one encoding function with an
additional input $W$: $f_{u,v}(Z,W):=f_{u,v,W}(Z)$. Therefore, by
adding $W$ to the input of each non-broadcast node, the resultant
network accepts $p_{X^{n},W}$. Refer to Figure \ref{fig:xor} for
the conditional XOR checker.

\medskip{}

\begin{rem}
The condition signal and message signal of a checker are analogous
to the ground pin (GND) and the power supply pin (VCC) of an IC chip
respectively. The ground is often connected to each internal component
of the chip, whereas the condition signal is connected to each non-broadcast
node of the checker. The ground is connected to zero volts by default,
whereas the condition signal is not connected to any message by default
(i.e., connected to ``zero information''), which reduces the conditional
checker to the original unconditional one. A chip works when there
is a voltage difference between the ground and the power supply (otherwise
it may lead to unexpected and undesirable outcomes), whereas a checker
works when there is a difference in information between the condition
signal and the message signals. If the message signals are not connected
(connected to ``zero information''), or if the condition signal
is connected to all messages (the highest amount of information),
the checker will always accept its inputs, and a gate will produce
meaningless outputs (that are any functions of the inputs). 

Nevertheless, unlike an IC chip where the inputs to the ground and
power supply pins are not supposed to change, we can choose a suitable
condition signal to a checker in order to ``selectively disable''
the effect of some messages. Also, a checker can have multiple message
signal inputs.
\end{rem}
\medskip{}

\subsection{Tristate Buffer Gate and Checker\label{subsec:tristate}}

\begin{figure}
\begin{centering}
\includegraphics[scale=0.92]{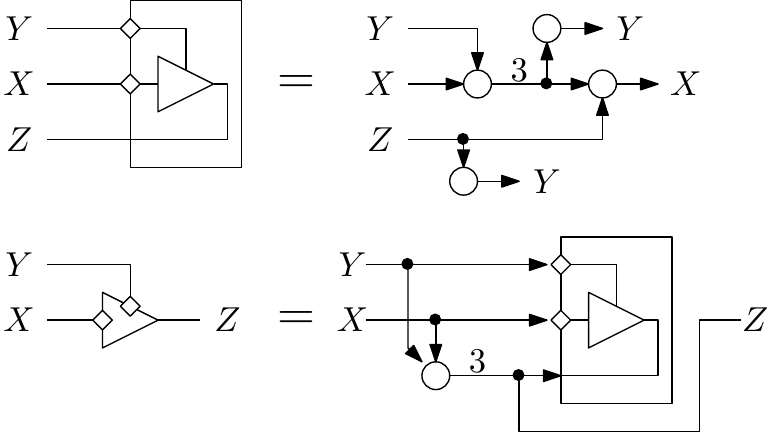}
\par\end{centering}
\caption{\label{fig:tristate}Top: The tristate buffer checker. Bottom: The
tristate buffer gate.}
\end{figure}

In digital circuits, the \emph{tristate buffer} has two inputs $X,Y\in\{0,1\}$,
and the output $Z\in[0..2]$ satisfies $Z=2$ (the high impedance
state) if $Y=0$, and $Z=X$ if $Y=1$. Assuming that $X$ is independent
of $Y$, we can check this by checking whether there exists $\tilde{Z}\in[0..2]$
such that
\[
H(Y|Z)=H(Y|\tilde{Z})=H(\tilde{Z}|X,Y)=H(X|Z,\tilde{Z})=0.
\]
To check this, note that $H(Y|Z)=0$ and $Z\in[0..2]$ implies that
$p_{Z|Y=y}$ is degenerate for at least one value of $y$. Same for
$p_{\tilde{Z}|Y=y}$. Since $H(X|Z,\tilde{Z})=0$, we know $p_{Z|Y=y}$
and $p_{\tilde{Z}|Y=y}$ must be nondegenerate for different values
of $y$. This implies that $p_{Z|Y=y}$ is nondegenerate for one value
of $y$. This gives us the desired distribution for $Z$. Refer to
Figure \ref{fig:tristate} for the construction of the tristate buffer
checker (assuming $X,Y$ are messages).

We can generalize this to a construction, which we call the $(b+1)$\emph{-state
buffer}, which has two inputs $X\in\{0,1\}$ and $Y\in[1..b]$, and
output $Z\in[0..b]$ with $Z=Y$ if $Y\ge2$, and $Z=X$ if $Y=1$.
Note that we use a different labelling of values compared to the tristate
buffer. Assuming that $X$ is independent of $Y$, we can check this
(up to relabelling) by checking whether there exists $Z_{2},\ldots,Z_{b}$
(let $Z_{1}=Z$) such that $Z_{i}\in[0..b]$ and $H(Z_{i}|X,Y)=0$
for $i\in[2..b]$, $H(Y|Z_{i})=0$ for $i\in[1..b]$, and
\begin{equation}
H(X|Z^{b})=0.\label{eq:bstate}
\end{equation}
We call the aforementioned conditions $\mathrm{BSTATE}_{b}(X,Y,Z)$.
The proof and the construction are similar to the tristate buffer.

\medskip{}

\subsection{Switch \label{subsec:switch}}

\begin{figure}
\begin{centering}
\includegraphics[scale=0.92]{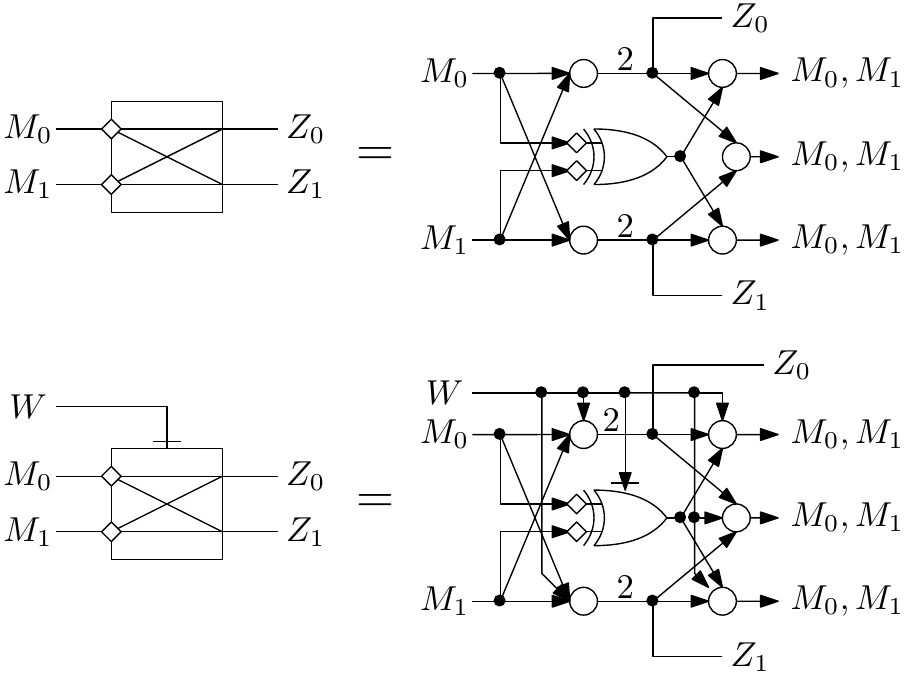}
\par\end{centering}
\caption{\label{fig:switch}Top: The switch gate. Bottom: The conditional switch
gate, which is constructed by adding $W$ to the input of every non-broadcast
node of the switch gate, and to the XOR gate (which is now the conditional
XOR gate). Note that unlike the XOR gate, the switch gate is not technically
a logic gate since the output not only depends on the inputs $M_{0},M_{1}$,
but also on the state $\theta$ (it is similar to a flip-flop, where
the output also depends on its internal state). In this paper, an
output of a gate does not need to be completely determined by its
inputs. We only require the outputs to satisfy some conditions depending
on the inputs.}
\end{figure}

In circuit-switching networks, a $2\times2$ crossbar switch is a
device with two inputs $M_{0},M_{1}\in\{0,1\}$, two outputs $Z_{0},Z_{1}\in\{0,1\}$,
and a state $\theta\in\{0,1\}$, where $(Z_{0},Z_{1})=(M_{0},M_{1})$
if $\theta=0$ (the ``bar'' state), and $(Z_{0},Z_{1})=(M_{1},M_{0})$
if $\theta=1$ (the ``cross'' state). Assuming $M_{0},M_{1}\stackrel{iid}{\sim}\mathrm{Unif}[0..1]$
are messages, we can check whether $(M_{0},M_{1},Z_{0},Z_{1})$ forms
a switch up to relabelling (for some fixed $\theta\in\{0,1\}$) by
checking whether
\begin{align}
 & H(M_{0},M_{1}|Z_{0},Z_{1})\nonumber \\
 & =H(M_{0},M_{1}|Z_{0},M_{0}\oplus M_{1})\nonumber \\
 & =H(M_{0},M_{1}|Z_{1},M_{0}\oplus M_{1})\nonumber \\
 & =0.\label{eq:switch}
\end{align}
To prove this, note that if the above condition holds, then we have
$H(Z_{0},Z_{1}|M_{0},M_{1})=0$ and $H(Z_{0},Z_{1})=2$, implying
$Z_{0},Z_{1}$ must be two distinct choices out of the following three
choices: $M_{0}$, $M_{1}$ and $M_{0}\oplus M_{1}$ (up to relabelling).
If $Z_{0}=M_{0}\oplus M_{1}$, then we cannot have $H(M_{0},M_{1}|Z_{0},M_{0}\oplus M_{1})=0$.
Hence we have $(Z_{0},Z_{1})=(M_{0},M_{1})$ or $(Z_{0},Z_{1})=(M_{1},M_{0})$.
Refer to Figure \ref{fig:switch} for an illustration of the switch
gate.

While the switch is functionally almost identical to the $2\times2$
crossbar switch in circuit-switching networks, for the purpose of
our construction, a more suitable analogy would be a bistable multivibrator
circuit which has two stable states (e.g. a flip-flop). Consider the
inputs $M_{0},M_{1}$ as two different ``voltages''. The outputs
$(Z_{0},Z_{1})$ can be either at the voltages $(M_{0},M_{1})$ or
$(M_{1},M_{0})$, and both states are ``stable'' (i.e., satisfy
the decoding requirements). Similar to how flip-flops can serve as
the basic memory units of a digital circuit, we also regard the state
$\theta\in\{0,1\}$ of the switch as the memory of our construction.

\begin{figure}
\begin{centering}
\includegraphics[scale=0.92]{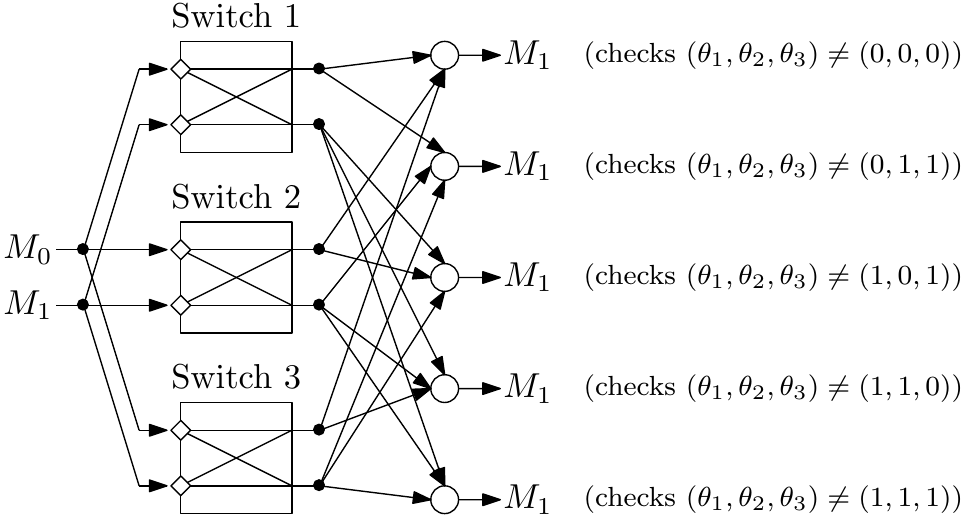}
\par\end{centering}
\caption{\label{fig:onehot}A set checker that checks whether $(\theta_{1},\ldots,\theta_{3})$
is a one-hot encoding.}
\end{figure}

 Consider an array of $n$ switches, with the same input $(M_{0},M_{1})$,
with outputs $(Z_{1,0},Z_{1,1}),\ldots,(Z_{n,0},Z_{n,1})$, and with
states $\theta_{1},\ldots,\theta_{n}$. We call this a \emph{physical
memory array} (to distinguish from the ``virtual memory array''
in the next section). We can impose the constraint that $(\theta_{1},\ldots,\theta_{n})\neq(a_{1},\ldots,a_{n})$
for any fixed $a_{1},\ldots,a_{n}\in\{0,1\}$ by requiring that $H(M_{1}|Z_{1,a_{1}},\ldots,Z_{n,a_{n}})=0$
(since $Z_{i,a_{i}}=M_{1}$ if $\theta_{i}\neq a_{i}$). By repeated
uses of this constraint, we can impose the constraint that $(\theta_{1},\ldots,\theta_{n})\in\Theta$
for an arbitrary set $\Theta\subseteq\{0,1\}^{n}$. We call a checker
that checks this constraint a\emph{ set checker }for\emph{ $\Theta$}.
Refer to Figure \ref{fig:onehot} for an example where we impose that
$(\theta_{1},\ldots,\theta_{3})$ is a one-hot encoding (exactly one
of them is $1$).

\medskip{}

\subsection{Conditional Switch and Virtual Memory Array \label{subsec:conditional_switch}}

Similar to the conditional XOR checker, we can also define the conditional
switch on $M_{0},M_{1},Z_{0},Z_{1},W$ (where $M_{0},M_{1}\stackrel{iid}{\sim}\mathrm{Unif}[0..1]$
is independent of $W$) by conditioning on $W$ on each term in \eqref{eq:switch}
(refer to Figure \ref{fig:switch} for an illustration). This allows
the state to depend on $W$, i.e., we have 
\[
(Z_{0},Z_{1})=(M_{\theta_{W}}\oplus\eta_{0,W},\,M_{1-\theta_{W}}\oplus\eta_{1,W}),
\]
where $\theta_{w},\eta_{0,w},\eta_{1,w}\in\{0,1\}$ for any $w\in\mathcal{W}$.
The terms $\eta_{0,w},\eta_{1,w}$ are needed since the switch checker
is only checking the distribution of $(M_{0},M_{1},Z_{0},Z_{1})$
up to relabelling, and hence in the conditional version of the switch
checker, the labelling for each $w$ may be different, and $Z_{0}$
may be flipped in different ways for different $w$. 

Therefore, one can create a \emph{virtual memory array} $\{\theta_{w}\}_{w\in\mathcal{W}}$
without physically adding more switches. The $w$-th bit $\theta_{w}$
can be ``retrieved'' by fixing the input $W=w$. This is analogous
to a memory array in digital circuit, with $W$ being the \emph{select
signal}. 

While a virtual memory array has the advantage of requiring fewer
switches (the number of nodes in the construction does not need to
scale with the number of bits in the memory), the operations that
can be performed are significantly more limited compared to a physical
memory array. For example, it is impossible to enforce that $\theta_{w}=0$
for a fixed $w$. One cannot even specify a value $w$ in a well-defined
manner since the network coding setting is invariant under relabelling
of the random variables. Therefore, any condition that can be enforced
on $\{\theta_{w}\}_{w\in\mathcal{W}}$ must be invariant under permutation
of the values in $\mathcal{W}$. 

\begin{figure}
\begin{centering}
\includegraphics[scale=0.92]{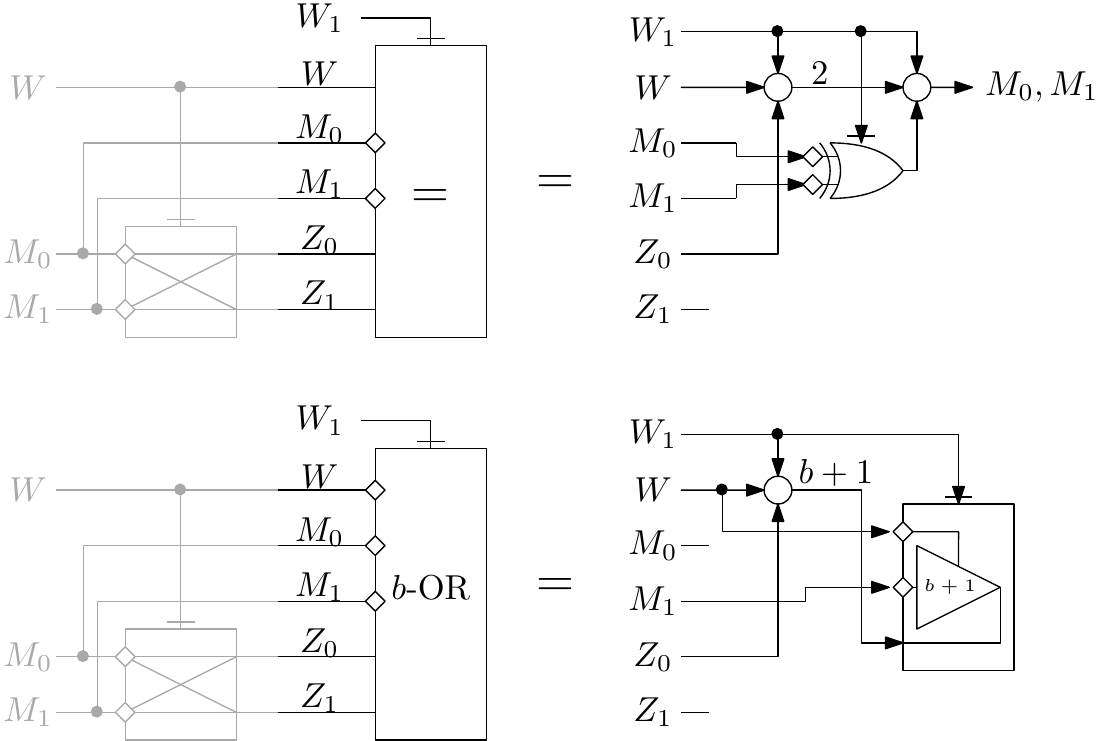}
\par\end{centering}
\caption{\label{fig:equal}Top: The conditional virtual equality checker with
select signal $W$, conditioned on $W_{1}$ (the conditional switch
on the left is included to demonstrate how the equality checker is
used; it is not a part of the equality checker). Bottom: The conditional
virtual $b$-ary OR checker, which uses the conditional $(b+1)$-state
buffer checker (note that $W$ must be a message, or a combination
of messages).}
\end{figure}

Assume $W\sim\mathrm{Unif}[1..b]$. We can check whether $\theta_{1}=\cdots=\theta_{b}$
by checking whether there exists $G\in\{0,1\}$ such that $H(G|Z_{0},W)=0$
and $H(M_{0},M_{1}|G,M_{0}\oplus M_{1})=0$. Note that $H(M_{0},M_{1}|G,M_{0}\oplus M_{1})=0$
implies that $G=M_{0}$ or $G=M_{1}$ (up to relabelling), which implies
$\theta_{1}=\cdots=\theta_{b}$ (otherwise it is impossible to determine
$M_{0}$ from $W$ and $Z_{0}=M_{\theta_{W}}\oplus\eta_{0,W}$, and
it is also impossible to determine $M_{1}$). For the other direction,
if $\theta_{1}=\cdots=\theta_{b}=0$, we can take $G=Z_{0}\oplus\eta_{0,W}$,
which equals $M_{0}$. The case for $\theta_{1}=\cdots=\theta_{b}=1$
is similar. Refer to Figure \ref{fig:equal} for the construction,
which we call the \emph{virtual equality checker} with select signal
$W$.

The \emph{conditional virtual equality checker} is the conditional
version of the virtual equality checker. Assume $W=(W_{1},W_{2})$,
where $W_{1}\sim\mathrm{Unif}[1..b_{1}]$ is independent of $W_{2}\sim\mathrm{Unif}[1..b_{2}]$.
The conditional virtual equality checker (conditioned on $W_{1}$)
is obtained by adding $W_{1}$ to the input of all non-broadcast nodes
of the virtual equality checker with select signal $W=(W_{1},W_{2})$.
It checks that $\theta_{w_{1},1}=\cdots=\theta_{w_{1},b_{2}}$ for
all $w_{1}\in[1..b_{1}]$. Refer to Figure \ref{fig:equal} for the
illustration.

Assume $W\sim\mathrm{Unif}[1..b]$. We can check whether $(\theta_{1},\ldots,\theta_{b})\neq(0,\ldots,0)$
by checking whether there exists $G\in[0..b]$ such that $H(G|Z_{0},W)=0$,
and the $(b+1)$-state buffer condition $\mathrm{BSTATE}_{b}(M_{1},W,G)$
\eqref{eq:bstate} holds. To prove this, note that if $\theta_{1}=1$,
then we can take $G=W$ if $W\ge2$, and $G=Z_{0}\oplus\eta_{0,1}=M_{1}$
if $W=1$. For the other direction, if $(\theta_{1},\ldots,\theta_{b})=(0,\ldots,0)$,
then we have $Z_{0}=M_{0}\oplus\eta_{0,W}$ independent of $M_{1}$,
so $G$ must be independent of $M_{1}$, and $\mathrm{BSTATE}_{b}(M_{1},W,G)$
cannot hold. Refer to Figure \ref{fig:equal} for the construction,
which we call the \emph{virtual $b$-ary OR checker} with select signal
$W$ (note that $W$ must be a message, or a combination of messages,
due to the construction of $(b+1)$-state buffer). 

The \emph{conditional virtual $b$-ary OR checker} is the conditional
version of the virtual $b$-ary OR checker. Assume $W=(W_{1},W_{2})$,
where $W_{1}\sim\mathrm{Unif}[1..b_{1}]$ is independent of $W_{2}\sim\mathrm{Unif}[1..b_{2}]$.
The conditional virtual $b_{2}$-ary OR checker (conditioned on $W_{1}$)
is obtained by adding $W_{1}$ to the input of all non-broadcast nodes
of the virtual $b_{2}$-ary OR checker with select signal $W=(W_{1},W_{2})$
\footnote{Note that it uses the $(b_{2}+1)$-state buffer instead of the $(b_{1}b_{2}+1)$-state
buffer. Although $W$ has alphabet size $b_{1}b_{2}$ which seems
to violate the requirement of $\mathrm{BSTATE}_{b_{2}}$, it does
not violate the conditional version of $\mathrm{BSTATE}_{b_{2}}$
conditioned on $W_{1}$ since the conditional cardinality of $W$
given $W_{1}$ is $b_{2}$.}. It checks that $(\theta_{w_{1},1},\ldots,\theta_{w_{1},b_{2}})\neq(0,\ldots,0)$
for all $w_{1}\in[1..b_{1}]$.

Note that it is possible to combine virtual and physical memory arrays
by having a physical array of conditional switches. We call this a
\emph{virtual-physical memory array}, which is an array of $n$ conditional
switches, with the same input $(M_{0},M_{1},W)$, with outputs $(Z_{1,0},Z_{1,1}),\ldots,(Z_{n,0},Z_{n,1})$,
and with states $\{\theta_{w,i}\}_{w\in\mathcal{W},\,i\in[n]}$. As
in the case of physical memory array, a \emph{conditional set checker}
for $\Theta$ (which is the conditional version of the set checker
where $W$ is added to the input of all non-broadcast nodes) checks
that $(\theta_{w,1},\ldots,\theta_{w,n})\in\Theta$ for all $w\in\mathcal{W}$,
where $\Theta\subseteq\{0,1\}^{n}$ is an arbitrary set. 

\medskip{}

\subsection{Cycles \label{subsec:cycles}}

Some of the constructions in Sections \ref{subsec:cycles} and \ref{subsec:2d}
are similar to those in \cite{li2021undecidability}, though we present
them here in a different way using the digital circuit analogy.

\begin{figure}
\begin{centering}
\includegraphics[scale=0.92]{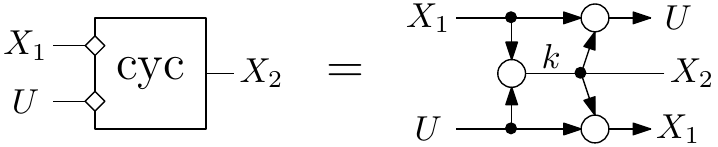}
\par\end{centering}
\caption{\label{fig:cycles}The cycles gate. The edge labelled $k$ has the
default size $k$ (i.e., $c_{E}(u,v)=0$).}
\end{figure}

In this section, we construct a network to check the cycles constraint
$\mathrm{CYCS}(X_{1},X_{2})$ in \cite{li2021undecidability}, which
is the constraint that $X_{1},X_{2}$ are uniform with the same alphabet
size, the pair $(X_{1},X_{2})$ is uniformly distributed over its
support, and all vertices in their characteristic bipartite graph\footnote{The characteristic bipartite graph is the graph with edge $(x_{1},x_{2})$
if and only if $p_{X_{1},X_{2}}(x_{1},x_{2})>0$.} have degree $2$, that is, the characteristic bipartite graph consists
of disjoint cycles. We also define $U\sim\mathrm{Unif}\{0,1\}$ as
in \cite{li2021undecidability}, which corresponds to the color in
a 2-coloring of the edges of the bipartite graph such that no two
edges sharing a vertex have the same color. Given $X_{1}\sim\mathrm{Unif}[0..k-1]$
(a default-size message) independent of $U\sim\mathrm{Unif}\{0,1\}$
(a fixed-size message), we can check whether $X_{2}$ satisfies the
cycles constraint by checking that $|\mathcal{X}_{2}|\le k$, and
\[
H(U|X_{1},X_{2})=H(X_{2}|X_{1},U)=H(X_{1}|X_{2},U)=0.
\]
It is straightforward to check that the above conditions are satisfied
if the cycles constraint is satisfied. For the other direction, assume
the above conditions are satisfied. Since $H(X_{1},U|X_{2},U)=0$,
$H(X_{1},U)=\log(2k)$ and $H(X_{2})\le\log k$, we have $U$ independent
of $X_{2}$. The rest follows from the same argument as in $\mathrm{CYCS}(X_{1},X_{2})$
in \cite{li2021undecidability}. Refer to Figure \ref{fig:cycles}
for the construction.

\medskip{}

\subsection{2D Memory Organization and Undecidability\label{subsec:2d}}

\begin{figure}
\begin{centering}
\includegraphics[scale=0.9]{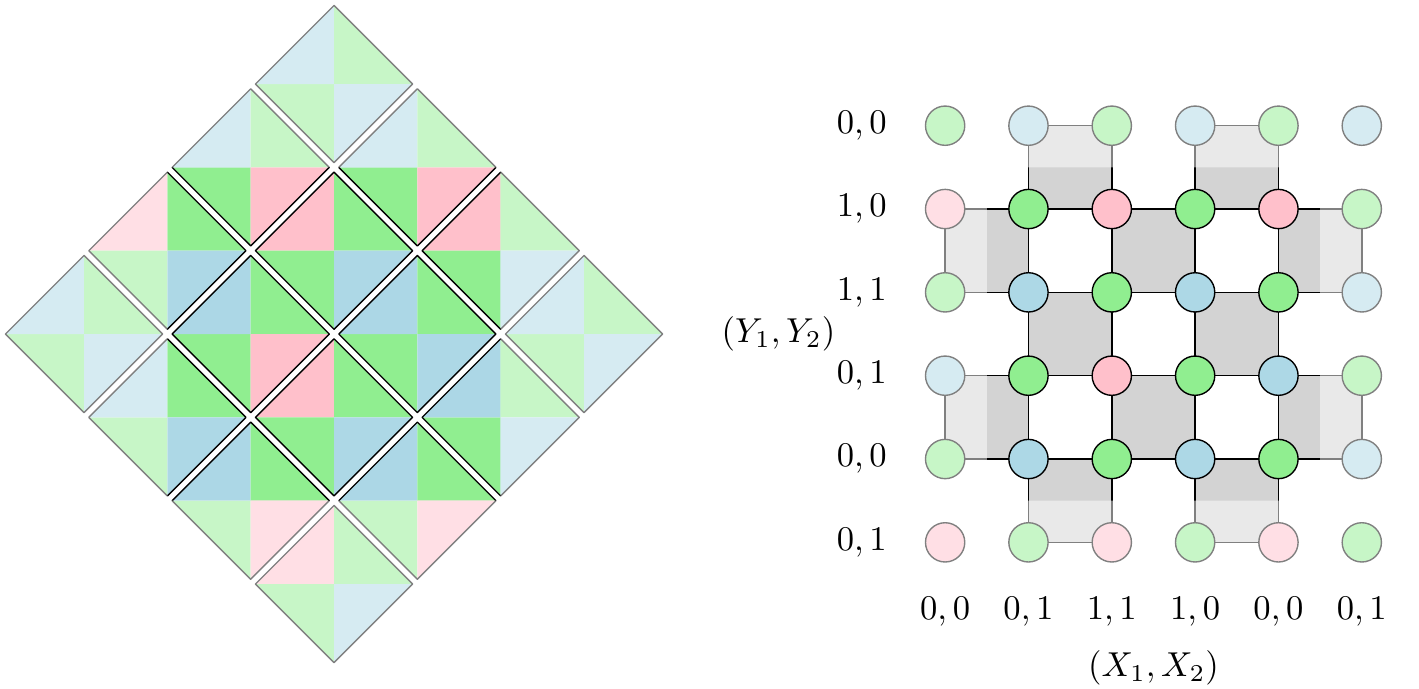}
\par\end{centering}
\caption{\label{fig:tiling}Left: A periodic tiling of a torus with $3$ colors,
with square tiles rotated $45^{\circ}$ (note that the top left and
the bottom right edges wrap around, and the top right and the bottom
left edges wrap around). Right: The corresponding coloring of the
$4\times4$ torus, where even faces are shaded. Each even face corresponds
to a tile. Each node $(x_{1},x_{2},y_{1},y_{2})$ in the torus has
a horizontal coordinate $(x_{1},x_{2})$ and a vertical coordinate
$(y_{1},y_{2})$. We will use a virtual-physical memory array with
a 2D memory organization to represent the colors of the vertices of
the torus.}
\end{figure}

In the periodic tiling problem \cite{wang1961proving,berger1966undecidability,gurevich1972remarks,mazoyer1999global},
we try to tile a torus with a set of square tiles, where each side
of the square is colored by one of $c$ colors. Given a set of tiles
(each specified by a $4$-tuple of colors), the problem is to decide
whether one can tile a torus using the set of tiles (repeated use
of the same tile is allowed, but no rotation or reflection is allowed),
such that adjacent tiles have the same color on their touching sides.
By treating sides of squares as vertices in a torus, we obtain the
grid of the torus (rotated $45^{\circ}$ compared to the grid of the
tiles). Therefore, the periodic tiling problem is equivalent to coloring
the vertices of a torus such that the 4 vertices in each even face
of the torus (like the black squares in a chess board) must have a
$4$-tuple of colors that is in the set of allowed $4$-tuples. Refer
to Figure \ref{fig:tiling} for an illustration.

The main idea of the proof of the reduction from the periodic tiling
problem in \cite{li2021undecidability} is to enforce the color restriction
of the torus using conditional independence relations and affine
existential information predicates (AEIPs) \cite{li2021automatedisit,li2021undecidability}.
Here we will follow the same general idea, but use gates and checkers
instead of AEIPs to enforce the color restriction. 

By taking two independent copies of cycles on $(X_{1},X_{2},U)$ and
$(Y_{1},Y_{2},V)$ (i.e., letting $X_{1},Y_{1}\sim\mathrm{Unif}[0..k-1]$,
$U,V\sim\mathrm{Unif}\{0,1\}$ all independent), we obtain a collection
of tori.  Each tuple $(x_{1},u,y_{1},v)$, or equivalently, a tuple
$(x_{1},x_{2},y_{1},y_{2})$ within the range of $(X_{1},X_{2},Y_{1},Y_{2})$
(recall that by the definition of cycles, one can deduce $U$ from
$X_{1},X_{2}$, and deduce $X_{2}$ from $X_{1},U$), corresponds
to a vertex in the collection of tori. Regard $(x_{1},x_{2})$ as
the horizontal coordinates of the vertex, and $(y_{1},y_{2})$ as
the vertical coordinates of the vertex. Following the definitions
in \cite{li2021undecidability}, two vertices $(x_{1},x_{2},y_{1},y_{2})$
and $(\tilde{x}_{1},\tilde{x}_{2},\tilde{y}_{1},\tilde{y}_{2})$ are
connected by a horizontal edge in the tori if $(x_{1},y_{1},y_{2})=(\tilde{x}_{1},\tilde{y}_{1},\tilde{y}_{2})$
or $(x_{2},y_{1},y_{2})=(\tilde{x}_{2},\tilde{y}_{1},\tilde{y}_{2})$.
Those vertices are connected by a vertical edge in the tori if $(x_{1},x_{2},y_{1})=(\tilde{x}_{1},\tilde{x}_{2},\tilde{y}_{1})$
or $(x_{1},x_{2},y_{2})=(\tilde{x}_{1},\tilde{x}_{2},\tilde{y}_{2})$.
We call a set of four vertices a type 11 face if their $x_{1}$ and
$y_{1}$ coordinates are the same. We call a set of four vertices
a type 22 face if their $x_{2}$ and $y_{2}$ coordinates are the
same. The type 11 faces and the type 22 faces are the even faces of
the tori. It was proved in \cite{li2021undecidability} that as long
as we can enforce the following three types of conditions, then the
color restriction in the periodic tiling problem can be enforced,
and hence the coloring problem of the tori (satisfying a list of such
conditions) is undecidable:

\smallskip{}

\begin{enumerate}
\item \textbf{Horizontal edge equality condition.} The colors $c_{1},c_{2}$
of the two vertices on each horizontal edge must satisfy $\mathbf{1}\{c_{1}\in\mathcal{C}\}=\mathbf{1}\{c_{2}\in\mathcal{C}\}$,
where $\mathcal{C}$ is any fixed set of colors. The vertical edge
equality condition is defined similarly.\footnote{This is used in the restriction $\mathrm{SAT}_{\neq1/2}$ in \cite{li2021undecidability}
that the sign of the colors within a torus must be the same.}\smallskip{}
\item \textbf{Horizontal edge OR condition.} The colors $c_{1},c_{2}$ of
the two vertices on each horizontal edge must satisfy $c_{1}\in\mathcal{C}$
or $c_{2}\in\mathcal{C}$, where $\mathcal{C}$ is any fixed set of
colors. The vertical edge OR condition is defined similarly.\footnote{This is used in the restriction $\mathrm{SAT}_{\le1/2}$ in \cite{li2021undecidability}
that each horizontal edge connect a group $1$ vertex and a group
$2$ vertex, or a group $3$ vertex and a group $4$ vertex; and that
each vertical edge connect a group $1$ vertex and a group $4$ vertex,
or a group $2$ vertex and a group $3$ vertex.}\smallskip{}
\item \textbf{Type 11 face OR condition.} The colors $c_{1},\ldots,c_{4}$
of the four vertices in each type 11 face must satisfy that $c_{i}\in\mathcal{C}$
for some $i\in[4]$, where $\mathcal{C}$ is any fixed set of colors.
The type 22 face OR condition is defined similarly. \footnote{This is used in the restriction $\mathrm{SAT}_{\le3/4}$ in \cite{li2021undecidability}
that enforces that each even face have colors that is in the set of
allowed $4$-tuples of colors.}\smallskip{}
\end{enumerate}
We now prove that these conditions can be enforced via gates and checkers.
Consider a virtual-physical memory array with $n$ conditional switches,
with $(X_{1},U,Y_{1},V)$ being the select signal. The memory of this
array is $\{\theta_{x_{1},u,y_{1},v,i}\}$ with $4nk^{2}$ bits. This
is analogous to the 2-dimensional memory organization in memory chips\footnote{Technically, it is closer to the ``2.5D'' organization which contains
a row decoder and a column decoder.}, with $(X_{1},U)$ being the column select signal, and $(Y_{1},V)$
being the row select signal. Let the set of colors be $[N]$. Each
color $c\in[N]$ is encoded into $n=2^{N}-2$ bits $\phi(c)\in\{0,1\}^{2^{N}-2}$,
where each entry of $\phi(c)$ is indexed by a nonempty proper subset
$\mathcal{A}\subsetneq[N]$, and $(\phi(c))_{\mathcal{A}}=\mathbf{1}\{c\in\mathcal{A}\}$.
We can therefore use the states of $n$ switches (indexed by subsets
of $[N]$) to represent a color. 

\begin{figure}
\begin{centering}
\includegraphics[scale=0.92]{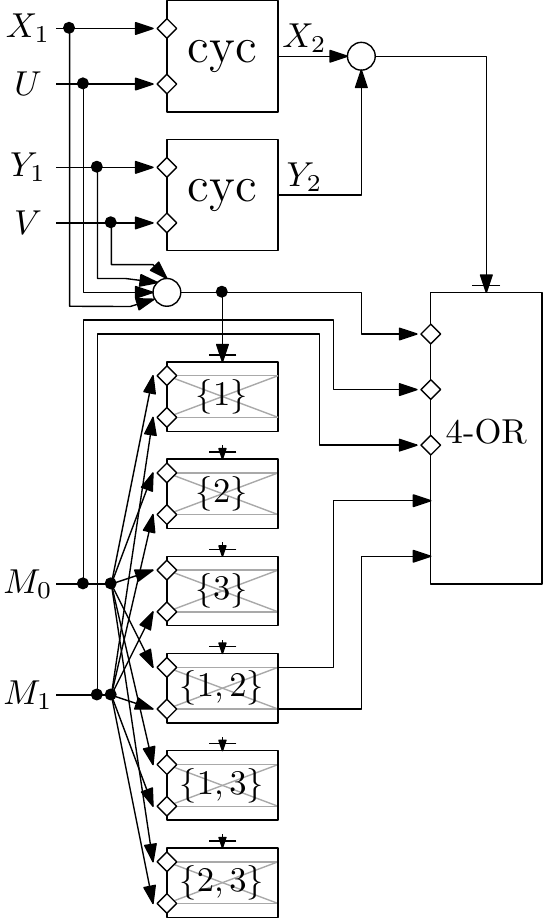}
\par\end{centering}
\caption{\label{fig:cycles-1}The 2D memory organization and a checker for
type 22 face OR condition, where the number of colors is $N=3$. The
switches are indexed by nonempty proper subsets of $[1..3]$. All
switches are conditioned on $(X_{1},U,Y_{1},V)$. The virtual $4$-ary
OR checker checks that every type 22 face cannot have four color-$3$
vertices (which is enforced if the all-color-$3$ tile is not in the
set of allowed tiles), or equivalently, at least one of the four vertices
has color in the set $\{1,2\}$. The other types of checkers are similar.
The conditional set checker for $\Theta=\{\phi(c):\,c\in[N]\}$ is
omitted.}
\end{figure}

Let the color of the vertex $(x_{1},u,y_{1},v)$ be $c(x_{1},u,y_{1},v)$.
We encode the collection of colors in the tori using the virtual-physical
memory array by $\theta_{x_{1},u,y_{1},v,\mathcal{A}}=(\phi(c(x_{1},u,y_{1},v)))_{\mathcal{A}}$.
We first need to enforce that $\{\theta_{x_{1},u,y_{1},v,i}\}$ is
a valid encoding. This can be enforced by a conditional set checker
(Section \ref{subsec:conditional_switch}) for $\Theta=\{\phi(c):\,c\in[N]\}$,
conditioned on $(X_{1},U,Y_{1},V)$. To enfoce the horizontal edge
equality condition for a set of colors $\mathcal{C}\subsetneq[N]$,
we use the conditional virtual equality checker (Section \ref{subsec:conditional_switch})
on the $\mathcal{C}$-th switch with select signal $(X_{1},U,Y_{1},V)$,
conditioned on $(X_{1},Y_{1},Y_{2})$, and another conditional virtual
equality checker conditioned on $(X_{2},Y_{1},Y_{2})$. This enforces
that any two nodes with the same $x_{1},y_{1},y_{2}$ (or $x_{2},y_{1},y_{2}$)
have colors $c_{1},c_{2}$ satisfying $(\phi(c_{1}))_{\mathcal{C}}=(\phi(c_{2}))_{\mathcal{C}}$,
i.e., $\mathbf{1}\{c_{1}\in\mathcal{C}\}=\mathbf{1}\{c_{2}\in\mathcal{C}\}$.
Similarly, to enfoce the horizontal edge OR condition, we use the
conditional virtual $2$-ary OR checker. To enfoce the face OR condition
for type 11 faces, we use the conditional virtual $4$-ary OR checker
with select signal $(X_{1},U,Y_{1},V)$, conditioned on $(X_{1},Y_{1})$.
To enfoce the face OR condition for type 22 faces, we use the conditional
virtual $4$-ary OR checker with select signal $(X_{1},U,Y_{1},V)$,
conditioned on $(X_{2},Y_{2})$. Note that we cannot use the select
signal $(X_{2},U,Y_{2},V)$ since the select signal of the virtual
OR checker must be messages ($X_{1}$ is a message, but $X_{2}$ is
not). Nevertheless, using $(X_{1},U,Y_{1},V)$ has the same effect
since they contain the same information as $(X_{2},U,Y_{2},V)$. Refer
to Figure \ref{fig:cycles-1} for an illustration of the checker for
type 22 face OR condition.

Recall that we assume all unlabelled edges in the diagrams have unlimited
sizes. It suffices to take the size of unlabelled edges to be the
product of the sizes of the messages in the network. If this exceeds
$k$, create enough parallel edges for each unlabelled edge (e.g.
if the product of sizes of messages is $10k^{2}$, then create three
parallel edges with sizes $10$, $k$ and $k$ for each unlabelled
edge). While technically the definition of the partially fixed-size
network does not allow parallel edges, we can create a relay node
for each parallel edge to make the edges distinct.

The proof is completed by invoking the reduction from the periodic
tiling problem to the problem of coloring the tori satisfying the
aforementioned three types of conditions proved in \cite{li2021undecidability}.

\medskip{}

\section{Index Coding}

We define one particular problem on index coding \cite{bar2011index,lubetzky2009nonlinear},
which we call \emph{partially fixed-size index coding}. The messages
are defined in the same manner as partially fixed-size network. There
are $l$ independent messages $M_{1},\ldots,M_{l}$. The alphabet
size of the message $M_{i}$ is $s_{M}(i)\in\mathbb{N}_{0}$, where
$s_{M}(i)=0$ means that $M_{i}$ has size $k$, where $k\in\mathbb{N}_{+}$
is the common default alphabet size. We have 
\[
M_{i}\sim\mathrm{Unif}[0\,..\,s_{M}(i)+k\cdot\mathbf{1}\{s_{M}(i)=0\}-1].
\]
The server observes $M^{l}$ and outputs $X=f(M^{l})$, where $f:\mathbb{N}_{0}^{l}\to[0..ak^{b}-1]$,
where $a\in\mathbb{N}_{+}$, $b\in\mathbb{N}_{0}$ are fixed (we enforces
that the alphabet size of $X$ is at most $ak^{b}$). There are $n$
clients, where client $j\in[n]$ observes $X$ and $M_{A_{j}}$, $A_{j}\subseteq[l]$,
and wants to decode $M_{B_{j}}$, $B_{j}\subseteq[l]$. The decoding
function of client $j$ is $g_{j}:\mathbb{N}_{0}^{1+|A_{j}|}\to\mathbb{N}_{0}^{|B_{j}|}$.
The decoding requirement is that $g_{j}(f(M^{l}),M_{A_{j}})=M_{B_{j}}$
for all $j\in[n]$ almost surely.

By the equivalence between network coding and index coding proved
in \cite{effros2015equivalence}, the partially fixed-size index coding
problem is also undecidable.
\begin{cor}
The following problem is undecidable: Given a partially fixed-size
index coding problem $(s_{M},a,b,\{A_{j}\},\{B_{j}\})$, decide whether
there exists $k\in\mathbb{N}_{+}$, encoding function $f$ and decoding
functions $\{g_{j}\}$ such that the decoding requirement is satisfied.
\end{cor}
\medskip{}

\section{Future Work}

We have proved that whether a given partially fixed-size network admits
a coding scheme is undecidable. By tracing the construction in the
proof, we can see that the only messages are $M_{0},M_{1},U,V$ (size
$2$) and $X_{1},Y_{1}$ (default size $k$), and only the sizes $2$,
$3$, $5$ and $k$ are required for edges. We conjecture that only
the sizes $2$ and $k$ are needed to prove undecidability.

\medskip{}

\begin{conjecture}
The following problem is undecidable: Given a partially fixed-size
network $(V,E,\{A_{v}\},\{B_{v}\},s_{M},s_{E})$ where $s_{M}(i),s_{E}(u,v)\in\{0,2\}$
for all $i,u,v$, decide whether it is solvable.
\end{conjecture}
\medskip{}
The main obstacle is to enforce the tristate and $(4+1)$-state buffer
conditions using only size-$2$ edges.

Another future direction is to study the asymptotic almost solvability
of the partially fixed-size network. In this setting, we have a default
size $k_{1}$ for default-size messages, and another default size
$k_{2}$ for default-size edges. The asymptotic capacity of the network
is 
\[
\underset{k_{2}\to\infty}{\lim\sup}\frac{\log k_{1}^{*}(k_{2})}{\log k_{2}},
\]
where $k_{1}^{*}(k_{2})$ is the largest possible $k_{1}$ for a fixed
$k_{2}$ such that the network admits a coding scheme. We conjecture
that finding the capacity is also undecidable.

\medskip{}

\begin{conjecture}
The following problem is undecidable: Given a partially fixed-size
network $(V,E,\{A_{v}\},\{B_{v}\},s_{M},s_{E})$, decide whether its
asymptotic capacity is at least $1$.
\end{conjecture}
\medskip{}
It was shown in \cite{lehman2005networkmodel} that allowing the rate
to be slightly below capacity can reduce the alphabet size drastically.
Therefore, it is unclear if the asymptotic almost solvability of the
partially fixed-size network is as hard as exact solvability.

\medskip{}

\section{Acknowledgement}

This work was supported in part by the Hong Kong Research Grant Council
Grant ECS No. CUHK 24205621, and the Direct Grant for Research, The
Chinese University of Hong Kong (Project ID: 4055133). The author
would like to thank Chandra Nair, Raymond W. Yeung, Andrei Romashchenko,
Alexander Shen, Milan Studený, Laszlo Csirmaz, Bruno Bauwens and Dariusz
Kaloci\'{n}ski for their invaluable comments.

\medskip{}

\[
\]

\bibliographystyle{IEEEtran}
\bibliography{ref}

\end{document}